# The physical construct of quantum mechanics

R. A. Street

Palo Alto Research Center, Palo Alto CA 94306

street@parc.com

**Abstract**

The physical constructs underlying the properties of quantum mechanics are explored. Arguments are given that the particle wave function as well as photon and phonon quanta must derive from a more fundamental physical construct that has not yet been identified. An approach to identifying the construct is discussed and a specific construct is proposed and explained. The proposal leads to a physical explanation of the wave function and quantized states, and a modified description of other quantum phenomena.

**Introduction**

Physics concerns the properties of physical constructs. The physical world is based on a set of elementary constructs, discovered empirically, with properties encapsulated in equations that allow an analysis of the physical phenomena. A goal of basic physics is to find the minimum essential set of fundamental constructs. Our understanding of the constructs has evolved over time with improved information and they were modified or replaced by others that are more general or better able to explain the measurements. Thus, protons and neutrons were found to be made up of quarks, and space-time replaced the Cartesian space of Newtonian gravity.

The conventional view is that quantum mechanics introduces no new fundamental physical constructs.[1] The rationale is that the analytical structure of Newtonian mechanics is shown by experiment to be insufficient to describe micro-scale properties and therefore has to be replaced by a new analytical structure. It is presumed to be sufficient to identify the correct analysis and that no new underlying physical construct is required. Hence, the discussion about the fundamentals of quantum mechanics mostly concerns postulates and axioms,[2,3] which are mathematical constructs not physical ones. The mathematical accounting of the difference between classical and quantum mechanics is that particle dynamics are described by vectors in a Hilbert space rather than by the classical phase space. Not only is quantum theory deep and elaborate, but its correctness has been widely verified and fully accepted. Its equations are routinely solved to explain measurements. Given the equations and a working interpretation of the solutions, then the rationale for not invoking a new physical construct seems reasonable and is supported by the fact that no obvious construct presents itself.

However, there are reasons to doubt that this view is correct. The most obvious and persuasive is that quantum theory introduces a new fundamental constant – Planck's constant $h$. The familiar commutation relation for position and momentum encapsulates the central role of Planck's constant in the difference between classical and quantum mechanics,

$$[\mathbf{p},\mathbf{x}] = 0 \quad \text{classical}$$
$$[\mathbf{p},\mathbf{x}] = i\hbar \quad \text{quantum}$$

The fundamental constants of physics are always associated with physical constructs, usually either a component of a fundamental particle or the strength of a force. A mathematical analysis cannot account for the specific value of $h$ – instead it is reflects the intrinsic magnitude of some physical construct. Hence the quantum mechanics properties must derive from a fundamental physical construct not identified in classical mechanics.

In addition, quantum mechanics does more than simply extend Newtonian mechanics. It introduces new aspects of physics not even hinted at classically, including particle wave functions and quantized states. It is actually not true that quantum mechanics introduces no new physical constructs because the electron (or any other particle) together with its wave function is a new physical construct, and is different from the classical construct of an electron as a point particle. The photon and phonon are, in the same way, also new constructs. However, it is unreasonable to argue that these constructs are fundamental because they exhibit essentially the same quantum properties despite being associated with completely different physical phenomena – particles, electromagnetic waves and mechanics vibrations. They each involve Planck's constant and some form of wave property. The de Broglie relation $\lambda = h/p$ for matter waves is essentially the same as the Planck relation $E = \hbar\omega$ for photons and phonons. An obvious explanation is that these and other aspects of quantum mechanics derive from a more elementary unifying physical construct.

As an aside, it is surprising how little the phonon has featured in the discussion about the physics and meaning of quantum mechanics. That two so completely different physical constructs, electromagnetic waves and mechanical oscillations, should obey the same quantization law, implies that their quantum physics cannot originate directly from either construct but instead its origin is from a more fundamental common factor.

Although quantum mechanics is successful, it comes with profound conceptual problems that have not been resolved in 100 years. These include the well-known issues relating to the superposition of states, the measurement problem, and the property that an apparently indivisible particle can be in two places at once, leading to questions about the nature of reality. Despite quantum theory being routinely applied without conceptual problems to a wide range of physical problems,[4] the conceptual problems have neither been resolved nor shown that they can be disregarded. The discussion of these problems is usually based on the mathematical theory but rarely on the physics. Several possible solutions to the various paradoxes have been proposed and analyzed without much consensus.[5] It is at least plausible that some additional knowledge about a physical construct underlying quantum mechanics might provide the basis for resolving the conceptual problems.

**Proposed identity of the quantum physical construct**

Based on these arguments, the hypothesis made here is that there is an underlying physical construct, whose intrinsic properties lead to quantum behavior including quantized states, the wave function and other properties. The immediate question is how to identify such a construct and what evidence is needed for it to be considered plausible. History shows that the discovery of a physical construct usually starts with empirical observations, from which the existence of a construct is deduced, its properties characterized and then encapsulated in equations. For quantum mechanics, the empirical observations and the equations are already known, but the hypothetical construct is unidentified. Therefore the requirement to identify the actual construct is that it qualitatively accounts for the empirical observations and plausibly is described by the equations of quantum mechanics. By definition, the quantum equations must describe the properties of the correct construct, if one exists, so it is sufficient that a candidate construct should have a clear and understandable relation to the relevant equations. The correct construct must provide a satisfactory qualitative explanation of wave functions, quantized states, measurement uncertainty, entangled states and so on. The mathematics of quantum mechanics started with the Schrödinger equation and developed to the Dirac equation, along with operator formalism and quantum field theory. However, it is sufficient for the construct to be described by the Schrödinger equation. The reason is that operator formalism and quantum field theory has added generalization, computational efficiency, deeper abstraction but no new physics. The Schrödinger equation remains valid and the physical construct it may describe is therefore automatically described by quantum field theory – only the mathematical formulation is different.

It may seem a hopeless task to identify the construct but the possibilities are actually few. Most elementary physical constructs are new particles or new forces. However, Dirac pointed out that there can be no new forces,[6] and no new particles are evident – given that the photon and phonon are properties of the unknown construct and not the construct itself. The most puzzling aspect of quantum mechanics is that it is classical mechanics with additional properties – it concerns position, mass, momentum, frequency and energy of a particle or collections of particles and waves of different types. As is apparent from Eq. 1, the rules of quantum mechanics do not suggest an obvious new construct.

This leads the search in a different direction, to ask if there a quantity that classical mechanics does not recognize as a physical construct but that actually is one. Discovering a new construct within the existing concepts of mechanics would lead to new properties ignored in classical mechanics but which plausibly provide the physics of quantum mechanics. This is not such an unusual idea as it might seem as there is a well-known example of such a paradigm shift. The Theory of Relativity replaced simple coordinate space with the new physical construct of space-time, imbued with physical properties – a mass-dependent curvature – not present in coordinate space. For quantum mechanics the obvious, and perhaps only, candidate for this type of construct is energy. Energy is conventionally thought of as a simple enumeration of a particular measureable quantity, but it is here proposed that it is a physical construct in its own right. Classical mechanics results when energy is treated as just a number while including the physical properties of energy leads to quantum mechanics. The transition from the quantum to the classical world occurs when the physical properties of energy, as opposed to its numerical value, have a negligible effect on a measurement.

**Evidence for the proposed new paradigm**

Energy is the proposed physical construct underlying quantum mechanics. Energy is unique amongst measurable quantities as being conserved across all interactions, and being equivalent to mass. It therefore has some general attributes that would be expected for a physical construct. Also quantum mechanics applies broadly to all aspects of physics (with its role in gravity undetermined) and the same is true of energy.

The evidence that energy could be the underlying physical construct of quantum mechanics comes from the early empirical development of the field. Planck showed that the $E=\hbar\omega$ relation was required to understand black body radiation but did not suggest an underlying construct to explain the new relation.[7] Since energy quantization is not contained in Maxwell's equations, one can suppose that the Planck relation is a property of the fundamental construct and by inspection it can be explained as a property of energy as a physical construct, in which the allowed energy of a wave depends on the frequency. This approach also immediately explains why phonons obey the same quantization law, since the law arises from the properties of energy and not directly from the specific nature of the wave. The de Broglie relation $\lambda=h/p$, describes a particle wavelength in terms of its momentum but as noted above it can be recast into the same form as the Planck relation. The photon, the phonon and the particle wave function therefore are unified by their common relation to the proposed energy construct.

The Schrödinger equation describes the energy of a particle, and it is normally understood as defining the allowed energy states of the particle, arising from the nature of the particle wave function. It is only a small adjustment to think of the equation as defining the energy construct, with the wave function being a property of the energy construct rather than of the particle. In this model, the energy construct and the particle are inextricably associated and the Schrödinger equation described this association.

Energy as a new physical construct therefore fits these early qualitative observations, and the primary characteristic of the energy construct is that is has wave-like properties and is hence extended in space. The energy construct qualitatively accounts for the empirical measurements and has a clear and understandable relation to the Schrödinger equation, hence meeting sufficient evidence to be considered further.

**A model for quantum properties**

In the proposed description of particle mechanics, the physical construct of energy and an energetic particle are inextricably linked although are fundamentally different constructs. They are linked because the system gains energy from a force on the particle, but the energy has its own physical properties which must be satisfied. The particle (e.g. an electron) is point-like but the associated energy is wave-like and extended. This model therefore explains wave-particle duality and the Principle of Complementarity, because the associated particle-energy state really does have wave and particle properties simultaneously. The different spatial properties of the components of the pair are the origin of the commutation relations. The uncertainty principle and the probabilistic outcome of a measurement arise because an experiment measures either the particle or the energy but not both. The aspect that is not measured is only known probabilistically through the linkage between the energy state and the particle. The model is consistent with the Copenhagen interpretation relating the wave function to the probability of finding the particle at a certain position.

In this model, energy quantization arises naturally, even predictably, from the need to satisfy simultaneously the different physical properties of the two associated constructs. The energy of a quantum system has to be consistent with both the classical energy of the particle or wave and with the additional physical properties of the energy construct. The classical component and the associated energy state in general have energy described by different functions of time and space, respectively $F_C$ and $F_E$, and the only physically allowed states are those when the two components have mutually consistent functions describing the allowed energy $E_A$,

$$E_A = F_C(\mathbf{x},t) \cap F_E(\mathbf{x},t) \qquad (2)$$

The allowed energies are the intersection of the set of energy solutions determined by the two functions. Hence even when each individual component has an energy described by a continuous function of space and time, allowed states may only occur at a finite set of discrete energy values, which are the quantized states. Thus, oscillations are quantized because the energy construct adds the relation $E=\hbar\omega$, as given by standard quantum theory, resulting in photons and phonons. Electron states of an atom are quantized because the Coulomb energy is only compatible with the energy construct at a few spatial configurations, those being described by the Schrödinger equation, which can be seen as an expression of Eq. 2. On the other hand a particle moving in free space is not quantized because any kinetic energy is allowed by the energy construct in this situation. In any equation of physics, the underlying physical constructs can be identified by the fundamental constants in the equation. The quantized electron energy levels of the atom contain two fundamental constants $\varepsilon_0$ and $m$, arising from the electron and $h$ arising from the energy construct.

Quantum mechanics applies to clusters of particles, a property that has caused some conceptual problems. The 2-slit matter diffraction measurements has been demonstrated in $C_{70}$ molecules,[8] and quantized nano-mechanical oscillations are also observed.[9] Virtually all phonons are the quantization of vibrations of multiple atoms, acoustic phonons involving an essentially infinite number. The wave function is evidently not a property of a single elementary particle but of the group. The proposed energy model explains this behavior. The concept of a particle with its associated energy state is readily extended to an atom or atom cluster with its energy state defining its dynamics.[10] The model provides a transition from the quantum to the macroscopic world.

As the size of a cluster increases, the spatial extent of the wave function decreases relative to the size of the cluster, becoming increasingly insignificant to the properties, so that eventually the measurable properties only reflect the magnitude of the energy and not its spatial properties.

The energy construct model does not resolve the conceptual problems of quantum mechanics but the proposed paradigm modifies the arguments and may ultimately help in their resolution. In the two-slit electron diffraction experiment, it has been argued that since the wave function passes through both slits then part of the electron must too,[11] in contradiction to the idea that the electron is an indivisible particle. In the proposed new paradigm, the wave function and the electron are different physical constructs. Hence it is reasonable to suppose that the diffracting energy wave goes through both slits while the electron goes through one or the other, although it is not known which one. A similar possibility applies to the superposition of multiple states which implies that the electron is in two states at once until resolved by wave function collapse. This type of paradox gets harder to understand when quantum properties are extended to the macroscopic world and has led to strange explanations.[12,13] The energy construct model attributes the wave function as a property of the energy state associated with the particle and there is no necessary implication that the physical objects themselves are simultaneously in two states. The energy construct may also provide some conceptual help in understanding entangled states. A pair of entangled electrons is described by a single wave function. The energy construct description implies that there is a real physical connection between the particles even when they are far apart, providing a physical mechanism for the observed correlation of properties.

**Discussion**

The conclusions reached in this paper rest on two main points, the first of which is that there are strong arguments to suggest an underlying physical construct from which quantum properties originate, and that the particle wave functions, photon and phonon quantization etc., derive their properties from this underlying construct. The second point is the proposal that energy is the physical construct. This model qualitatively accounts for the key experimental observations and the Schrödinger equation can be understood to describe its properties, ensuring consistency with the mathematical development of quantum theory. Both of these aspects of the paper are outside of the conventional view which is that no underlying construct is needed for quantum mechanics and that energy is a simple numerical value. The arguments given above for the existence of an underlying construct are the key point, because they open up the prospect of fully describing the construct. It is possible that there is an even more fundamental construct that the proposed energy state.

Given that such a physical construct exists, the proposal that it is energy raises the questions of whether this model can be proven, what exactly is the energy construct and what difference does it make to identify the construct? A pertinent question is whether it is a provable model. The reason to doubt its provability is that quantum mechanics has developed into a full working theory without needing the construct. However, the proposal that energy is a physical construct with its own properties and is not a purely numerical value, is surely either right or wrong. It may not be proved by the evidence presented here but it is not unprovable.

The properties attributable to energy as a real physical construct are that it is extended in space, with wave-like properties that depend on frequency and the magnitude of the energy. The lack of further detail seems to demand more information about what the energy state is. However, other physical constructs are no different since, for example, there is no good answer to the question, what is an electron made of? Fundamental physical constructs are defined by their properties; the electron is defined by its mass, charge, spin and size. The properties of the energy construct are known because they are described by the equations of quantum mechanics.

There are several answers to the question of what difference does it make to identify energy as the underlying construct and one is the satisfaction of having an answer. More important, the concept allows other physical phenomena to be examined in a new light. The Dirac equation extends the quantum theory to relativistic conditions, and the key result is the discovery of negative energy states, and their association with anti-particles. Negative energy is obviously not compatible with the conventional view of energy as a strictly positive number. The initial interpretation accepted negative energy at face value and supposed that all the negative energy states were filled and a positron, for example, arose when an excitation left one of the filled electron states unoccupied. This supposition raised the conceptual problem of having an infinite number of negative energy states filled by an infinite number of charged particles. Later came the currently preferred idea that the negative energy states represent a particle going backwards in time,[14] also not an intuitive idea. The energy construct model provides a different approach. As a physical construct with physical properties, rather than a numerical value, energy can have positive and negative values, just as there is positive and negative electrical charge. It may be a property of the construct that negative energy associated with an electron imbues it with the properties of a positron.

A different approach to finding evidence for the energy model is to consider the possible implications beyond quantum mechanics. In the physical situations described above, the energy state is associated with another physical entity, which raises the question of whether the energy construct can exist independently. The obvious candidate is cosmological dark energy, suggesting the possibility that dark energy is a form of energy that is not associated with any particles or other excitation and, being spatially extended, plausibly has a component of pressure by virtue of freely expanding in the universe. The density of dark energy is similar in order of magnitude to that of matter and radiation, and suggests a related origin. Models relate the photon redshift to the work done by radiation pressure on the expanding universe, but the mechanism for the transfer of energy is unclear. A possibility is that this transfer of energy to the universe releases physical energy from electromagnetic waves and creates dark energy. This is an obviously speculative idea, but shows that the proposal of the energy construct can lead to new lines of thought to explore.

**Summary**

In summary, the introduction of a new fundamental constant, and the fact that the same quantum relation applies to matter, electromagnetic waves and mechanical oscillations, are arguments that there is an underlying physical construct of quantum mechanics. The paper goes on to propose that energy is the missing physical construct and that its wave-like physical properties are the origin of quantum mechanics. The properties of energy combine with the classical properties of particles and oscillatory excitations to give the complete description of the various quantum mechanical systems. According to this model, energy has the property of being spatially extended, and the wave function reflects its actual spatial extent. Attributing quantum mechanics as laws describing the property of energy does not change the equations or their solutions, but does account for the physical origin of the properties.

• Wave-particle duality of matter particles is explained by the system actually comprising two associated physical components one of which is a particle and the other an energy wave. Duality of electromagnetic waves arises because there is an indivisible particle-like minimum energy due to the physical nature of energy.

• Quantized energy levels are explained as arising from the need to satisfy simultaneously the two different constraints of the associated physical entities.

• The equations of quantum mechanical and their formulation as quantum field theory are understood as describing the physical properties of energy as they apply to particles or oscillations.

• The collective quantum properties of atom clusters are explained by attributing their properties to the cluster energy rather than to the individual particles.

**Acknowledgements**

The author is grateful for helpful discussions with D. Biegelsen, W. Jackson and D. Tong.